# Applications of Sequential Learning for Medical Image Classification

Sohaib Naim[1], Brian Caffo[2], Haris I Sair[3,4], Craig K Jones[3,4,5]

[1]Department of Biomedical Engineering, Johns Hopkins University, Baltimore MD
[2]School of Public Health, Johns Hopkins School of Medicine, Baltimore MD
[3]Department of Radiology and Radiological Science, Johns Hopkins School of Medicine, Baltimore MD
[4]Malone Center for Engineering in Healthcare, Johns Hopkins University, Baltimore MD
[5]Department of Computer Science, Johns Hopkins University, Baltimore MD

Corresponding Author:

Craig Jones, PhD
Malone Hall, Suite 340
3400 North Charles Street
Baltimore, MD 21218-2608
craigj@jhu.edu

# Abstract

Purpose: The aim of this work is to develop a neural network training framework for continual training of small amounts of medical imaging data and create heuristics to assess training in the absence of a hold-out validation or test set.

Materials and Methods: We formulated a retrospective sequential learning approach that would train and consistently update a model on mini-batches of medical images over time. We address problems that impede sequential learning such as overfitting, catastrophic forgetting, and concept drift through PyTorch convolutional neural networks (CNN) and publicly available Medical MNIST and NIH Chest X-Ray imaging datasets. We begin by comparing two methods for a sequentially trained CNN with and without base pre-training. We then transition to two methods of unique training and validation data recruitment to estimate full information extraction without overfitting. Lastly, we consider an example of real-life data that shows how our approach would see mainstream research implementation.

Results: For the first experiment, both approaches successfully reach a ~95% accuracy threshold, although the short pre-training step enables sequential accuracy to plateau in fewer steps. The second experiment comparing two methods showed better performance with the second method which crosses the ~90% accuracy threshold much sooner. The final experiment showed a slight advantage with a pre-training step that allows the CNN to cross ~60% threshold much sooner than without pre-training.

Conclusion: We have displayed sequential learning as a serviceable multi-classification technique statistically comparable to traditional CNNs that can acquire data in small increments feasible for clinically realistic scenarios.

## Section 1: Introduction

Numerous improvements to image processing hardware and software have advanced radiology through computer vision (CV) and artificial intelligence (AI) applications [1]. A subset of AI involved with radiology research is the use of machine learning (ML) to better understand medical imaging data. Modern ML approaches have seen major challenges which limits its usage in standard radiology practices. Traditional neural networks for image classification utilize large data repositories split into training, validation, and testing subsets [2] though such networks are also limited by many training epochs, potentially long run-times, and considerable computational power. Additionally, the cost of collecting and annotating retrospective data is expensive. Day to day, radiologists observe a variety of cases during their regular clinical work and this process could be leveraged to train a neural network regularly. However, the limited training data during early stages of a new clinical study present challenges when translating traditional ML techniques to real-world medical imaging cases. The level of effort and expenses make developing commercial models non-viable due to difficulty establishing the dataset that would form the basis of training [3].

Sequential learning is an ML approach that trains a neural network with few datasets at a time. This technique is also referred to as incremental learning which provides the ability to maintain life-long learning for multiple classes' data. Incremental learning also maintains good performance on present and future data streams and requires shorter runtimes to update the network, which is advantageous under constrained resources [4]. As the metric for time to train a neural network continues, learned characteristics of initial class information can be forgotten in place for newer class information, a phenomenon known as catastrophic forgetting [4]. Newer datasets of the initial classes are also capable of impacting the distribution of all the training data the network has been learning from. This phenomenon, referred to as concept drift, occurs when enough key differences between initial and future datasets can negatively degrade the performance of the classification network [4]. In this work we will explore the deficiencies caused by these two obstacles as well as reducing likelihood of overfitting. As a neural network is trained on specified data, the training accuracy would increase at a steady rate. However, fluctuations due to minor details and noise could be misinterpreted by the network as important features. This may result in a model which memorizes all data instead of learning the discipline hidden behind the data, severely hindering the model's ability to generalize on future oncoming datasets [5].

Sequential learning is related to incremental learning, progressive learning, online learning, and lifelong learning. Though the subtleties with each technique are mostly opinion-based from experts within the research spectrum, they are distinguished by their ability to adapt to changes in data distribution for continuous training in real-time. Incremental learning refers to data streaming strategies which work under limited memory resources [6]. Sequential learning learns a sequence of tasks without having access to training data from previous or future tasks [7]. Progressive learning, while similar to other continual training techniques, has its information fixed on a number of classes and is ultimately restricted from learner newer classes on the run [8]. Online learning algorithms can adapt to newer datasets and are preferable for large-scale datasets and real-time problems [9]. However, often due to limited resources they are vulnerable to forgetting previously learned information through catastrophic forgetting and concept drift [6,9]. Lifelong learning is a form of continual learning that can learn streams of information, where the information is steadily made available over time and the number of tasks are not predefined [10]. Catastrophic forgetting is also mitigated with this approach by repeatedly showing the network the same pseudo-randomly shuffled training examples to recover lost knowledge [10].

Different methodologies have been proposed to address the limitations of continual learning techniques. For implementation in clinical practice, ML models should be capable of dealing with a continuous data stream from different sources. Perkonigg et al. displayed this method, known as dynamic memory (DM), using cardiac magnetic resonance imaging (MRI) segmentations acquired from four different vendors and computer tomography (CT) lung nodule data acquired from the Lung Image Database Consortium [11]. Performance degradation is another limitation that afflicts continual learning due to data distribution shifts over time. Srivastava et al. formulated a domain incremental learning approach with large-scale chest x-ray datasets with clear domain shifts [12]. Such shifts can be attributed to changes in acquisition hardware which can severely impact the performance of any modern ML model. They experimented on three public medical imaging datasets, one of which being the National Institutes of Health (NIH) Chest X-Ray Dataset. They utilized minimal memory space while also proving to be more resilient against catastrophic forgetting compared to what others have shown. Zhu et al. addresses the problem of overfitting through incremental learning with new classes but with fewer samples [13]. They addressed problems such as expanding initial representation space and retaining old class knowledge when introducing new class data with their incremental prototype learning scheme. They showed an increased number of iterations resulted in a corresponding accuracy decrease, likely due to a direct relationship with training difficulty and iteration count. These results substantiate our objective to find an optimal number of training iterations for a sequentially trained network.

In the following sections we discuss our sequential learning implementation and describe two findings. The first finding is to compare sequential learning with and without initial pre-training. We predict comparable performances between sequential learning methods with each presenting their own advantages. The second finding is to compare two methods of organizing a validation dataset to mitigate over-fitting during sequential learning. We predict the method which acquires training data across multiple "days" would best mitigate over-fitting. Lastly, we will define future directions for our sequential learning pipeline.

## Section 2: Materials and Methods

### 2.1 Datasets

All experiments were conducted using medical imaging datasets which have been made publicly available on Kaggle, the first of which was the Medical MNIST dataset contributed by Arturo Polanco Lozano [14]. This dataset consists of 58,954 JPEG images with size 64x64 across six imaging modalities [14]. A sample medical image from each modality can be seen below in the top row of Figure 1. The second dataset took advantage of the Medical MNIST dataset's CXR modality by applying data augmentation to rotate it 90 degrees to the left or the right. Each original CXR image would be selected at random to rotate in either direction. This provided three subclasses: Original CXR image, CXR image rotated 90 degrees left, and CXR image rotated 90 degrees right for a total of 10,000 JPEG images, same as the original Medical MNIST CXR class. The images were generated using shell script via the command line and maintained their dimensions of 64x64. A sample medical image of each class can be seen below in the middle row of Figure 1. The third dataset used was publicly available on Kaggle, known as the NIH Chest X-Ray dataset [15]. It consists of 112,120 PNG images in 15 classes each with 1024x1024 dimensions [15]. A sample medical image for each diagnosis can be seen below in the bottom row of Figure 1.

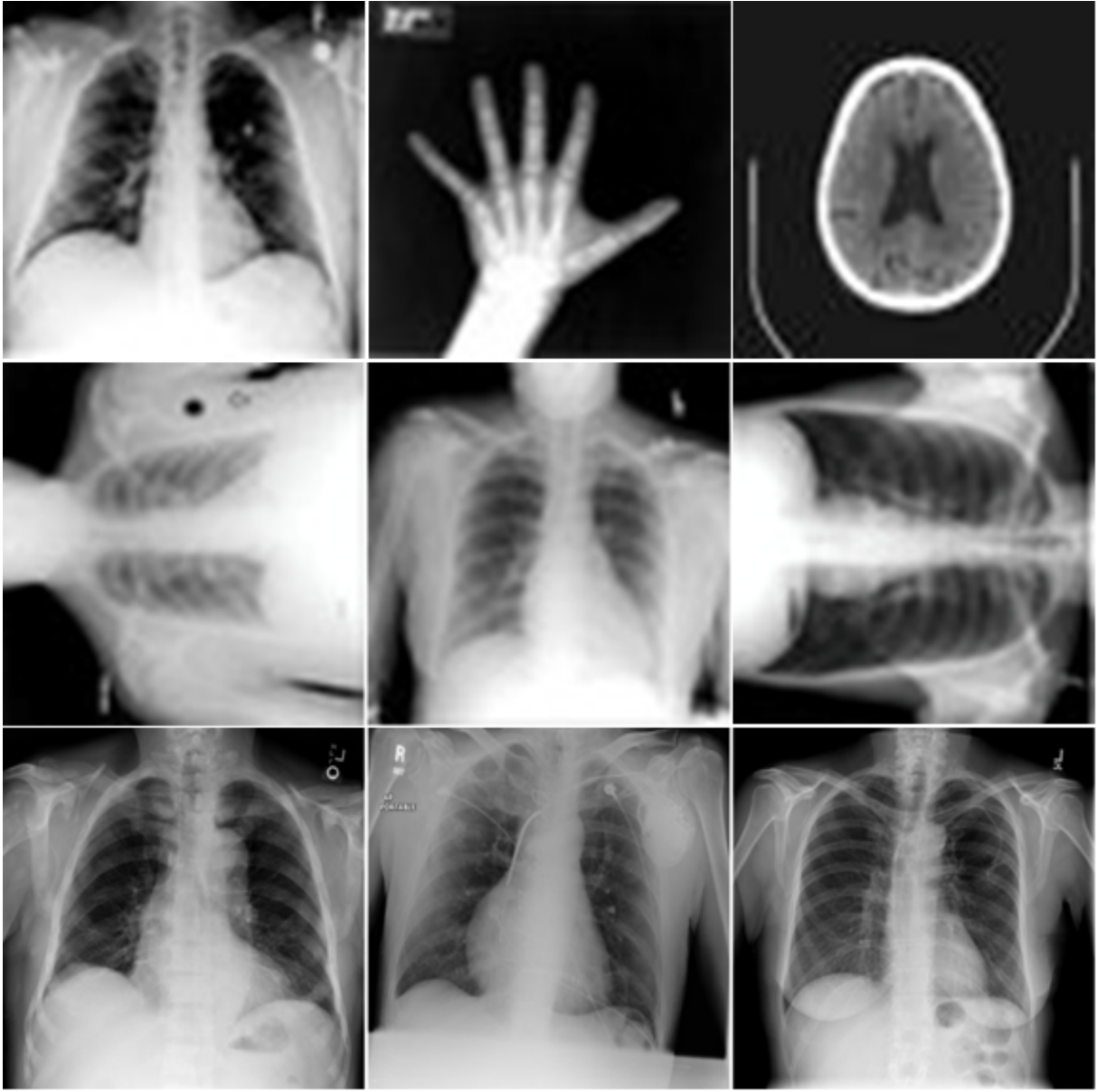

**Figure 1**: Sample cases for classes used in each dataset. Top row from left to right: MedicalMNIST CXR, Hand, HeadCT. Middle row from left to right: MedicalMNIST rotated left, original, rotated right. Bottom row from left to right: Kaggle Chest X-Ray data with effusion, mass, no finding.

## 2.2 Experimental Methods

In this work we will discuss our sequential learning implementation across three experiments. Experiment 1 represents the baseline experiments where we compare sequentially trained CNNs to evaluate the advantage of adding a pre-training component. Experiment 1 will use the three classes of CXR, Hand, and HeadCT from the Medical MNIST dataset. Experiment 2 will evaluate two methods of generating a validation dataset appropriate for a sequentially trained CNN and will use the three classes from our modified (rotated) CXR dataset. Experiment 3 compares two additional sequentially trained CNNs with and without pre-training, however this will be accomplished using the more challenging, and real-world, NIH Chest X-Ray dataset. Figure 2 shows the sequential learning experimentation with and without pre-training in Experiments 1 and 3. Figure 3 shows the validation dataset generation in Experiment 2.

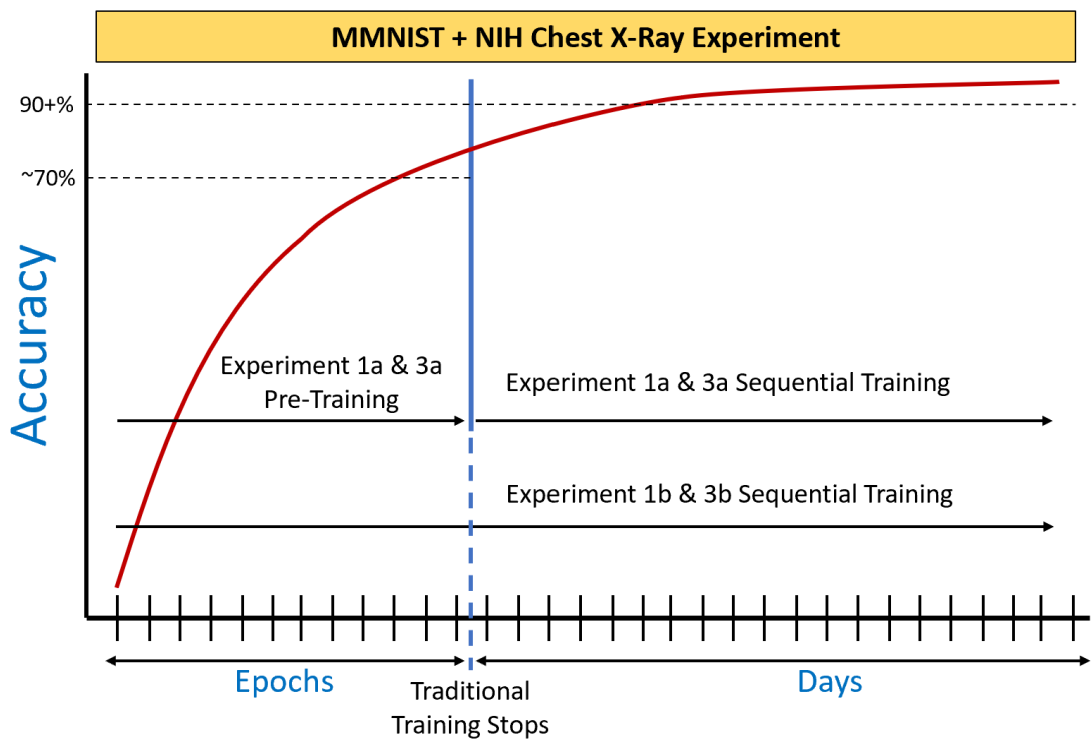

**Figure 2:** The MMNIST & NIH Chest X-Ray Experiment framework for CNN learning that includes batch pre-training followed by sequential learning. Experiment 1a & 3a includes a traditional training method with 3-way classification on an allocated set of training data before shifting to sequential learning for a defined number of datasets acquired per "day". Experiment 1b & 3b is performed with 3-way classification in a sequential learning format for a defined number of datasets acquired per "day".

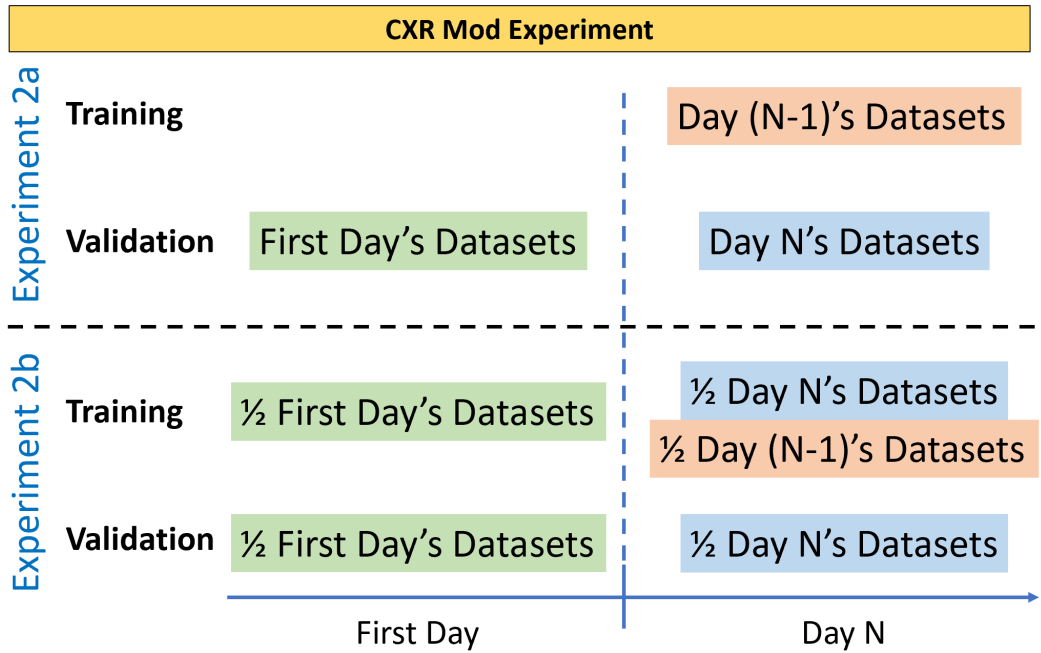

**Figure 3:** Breakdown of experiments that establish training and validation datasets used in a given day. For the first day, only the present day's data is used for training and validation splits. For day N, which represents any day that follows the first day, both the previous day and present day's combined data is used for the defined training and validation splits.

### 2.2.1 Experiment 1: Pre-Training vs No Pre-Training

The first experiment that would test our algorithm is a 3-way classifier on the Medical MNIST dataset. For each CXR, Hand, and HeadCT class there was an even split of 10,000 images for a total of 30,000 images split into training (70%), validation (10%), and testing (20%) subsets. The data splits were created once and written to text files to assure the same training, validation, and testing images were used for the experiment to maintain consistent findings. Experiment 1a included a pre-training loop, based on a separate subset of global training data, that was run until an accuracy of ~70% was obtained. Following pre-training on 500 images, the model would undergo validation against the global validation data. Following pre-training, the model was trained through sequential learning over a period of "days" and images per "day" until the sequential loss stabilizes at a minimum loss and sequential accuracy stabilizes to 90+%. The model will be evaluated against the test dataset after each training "day" to understand performance. Experiment 1b will use the same hyperparameters as Experiment 1a but will instead train through sequential learning exclusively over an increased number of training "days". Increasing the total number of "days" in Experiment 1b compensates for lack of pre-training to ultimately train on a comparable number of images as Experiment 1a.

### 2.2.2 Experiment 2: Validation Set Construction

In Experiment 2 we compared two methods of constructing a validation dataset based solely on the day-to-day data. The rotated CXR data was split into training (70%), validation (10%), and testing (20%) sets. The split was done once and these were used for consistency across all training in Experiment 2. For Experiment 2a, the N images arriving in the "current day" would be used as validation data, and the N images that arrived the "previous day" would be used as training data for the model. On the first "day", the CNN begins with the validation data using the untrained model as there is no "previous day's" training data. For Experiment 2b, N/2 images from the "current day" along with N/2 images from the "previous day" to be used as training data for the model, whereas the remaining N/2 images from the "current day" would be used as validation data. Unlike the previous experiment, there is training performed on the first "day" due to N/2 images arriving from the "present day". Following each "day's" training for both experiments, the CNNs were tested against the global testing data.

### 2.2.3 Experiment 3: Real-World Data

Experiment 3 used the NIH Chest X-Ray dataset [15] for the sequential learning as this was a more realistic "real-world" dataset that could better challenge our classification algorithm. For this experiment, we built a 3-way classifier using 2,100 images from each of the following classes: Effusion, Mass, and No Finding. These images were randomized and split into training, validation, and testing subsets with the above percentages. Experiment 3a includes the pre-training component on an allocated set of 1000 images from the global training data and is validated on the pre-training results against the global validation data. After training over a limited number of epochs, we expect to reach a desirable accuracy threshold (~50%) before shifting to a sequential learning format where the network will continue to train over a defined number of "days" and images per "day". After evaluating the model against the global testing dataset following each "day", we expect the sequential testing accuracy to stabilize between 65-75% as training nears completion. Experiment 3b will use the same hyperparameters as 3a but will undergo sequential training exclusively over an increased number of training "days". Like Experiment 1a, the increased number of "days" is to show a comparable number of images being used between experiments.

2.3 Neural Network Training and Statistics

We utilized the ResNet50 PyTorch model for Experiments 1 & 2 and a DenseNet121 PyTorch model for Experiment 3 as it provided better results for the NIH Chest X-Ray dataset shown by Tang et al. [16]. As stated above, the data splits remained consistent across all experiments. Data augmentation consisted of a random horizontal flip, +/- 5º rotation, horizontal and vertical translation (0.05% of the image), and color jitter (5%). We also normalize the images with mean and standard deviation values provided by ImageNet. Experiments 1 and 2 used a cross-entropy loss and Adam optimizer, whereas Experiment 3 used BCE with logits loss and an SGD optimizer. All hyperparameters are shown in table 1. Hyperparameters were optimized by comprising the learning rate, epochs and batch size. Accuracy and loss curves were used to evaluate the performance. All processing was done in PyTorch.

| PyTorch Hyperparameters | Experiment 1 | Experiment 2 | Experiment 3 |
|---|---|---|---|
| **Batch Size** | 16 | 16 | 64 |
| **Learning Rate** | 1e-6 | 1e-5 | 1e-3 |
| **Pre-training Epochs** | 5 | None | 5 |
| **Total Days** | 300 (1a), 500 (1b) | 10 | 170 (3a), 220 (3b) |
| **N datasets per day** | 20 | 50 | 20 |
| **N-epochs per day** | 1 | 150 | 5 |

**Table 1**: Hyperparameters used for each experiment.

## Section 3: Results

There were different experimental frameworks used between the three datasets. Beginning with the Medical MNIST dataset, we performed our baseline experiment to evaluate a sequential learning network with a pre-training component to a network without any pre-training and compared the results to ensure consistent findings. Both networks would train a multi-classification model with a single day-epoch over a long period of learning "days". The second experiment performed with the CXR dataset evaluated distinctive validation data recruitment methods over multiple day-epochs with fewer overall "days". The final experiment relates back to the first experiment evaluating pre-training against no pre-training. However, here we are using the more challenging NIH Chest X-Ray dataset as our "real-world" example.

## 3.1 Experiment 1: Pre-Training vs No Pre-Training

The results shown in Figure 4 represent the sequentially trained CNN with and without pre-training. Experiment 1a includes batch pre-training until a desirable starting accuracy was obtained by the network before it continues to train via sequential learning. We can observe how the testing accuracy begins at ~75%, displaying proper continuation of training. The model fluctuates between 60-95% accuracy over the first 100 "days" before stabilizing between 80-100% accuracy over the final 200 "days". There are a few precipitous dips that are observed, but the model quickly self-corrects shortly after and restabilizes. The testing accuracy increases steadily over the training period and stabilizes at ~95% for the final 100 "days". The testing loss steadily decreases and appears to stabilize at ~0.4, indicating full information extraction has been reached. Experiment 1b represents the CNN that begins sequential training from scratch over a period of 500 "days". The lack of pre-training is distinguishable by the testing accuracy that begins at ~45%. Training accuracy fluctuates between 20-60% initially due to the model being untrained. As the training period continues though the training accuracy increases with fewer fluctuations, eventually reaching 80-100% by "day" 250 and beyond. Testing accuracy similarly learns slowly early on but steadily increases to ~95% accuracy by "day" 500. As information is collected over the training period, the testing loss steadily decreases through "day" 500 despite not appearing to flatten, indicating that more training "days" should be considered. However, the testing accuracy in both experiments mostly stabilizes to ~95% and the testing loss reaches ~0.4, even if the latter is not fully stabilized by the end of the training period. These comparative metrics show how sequential learning remains a viable option despite lacking initial training data. The CNN would still benefit from initial training data though as it would require fewer timepoints to reach a desirable accuracy threshold.

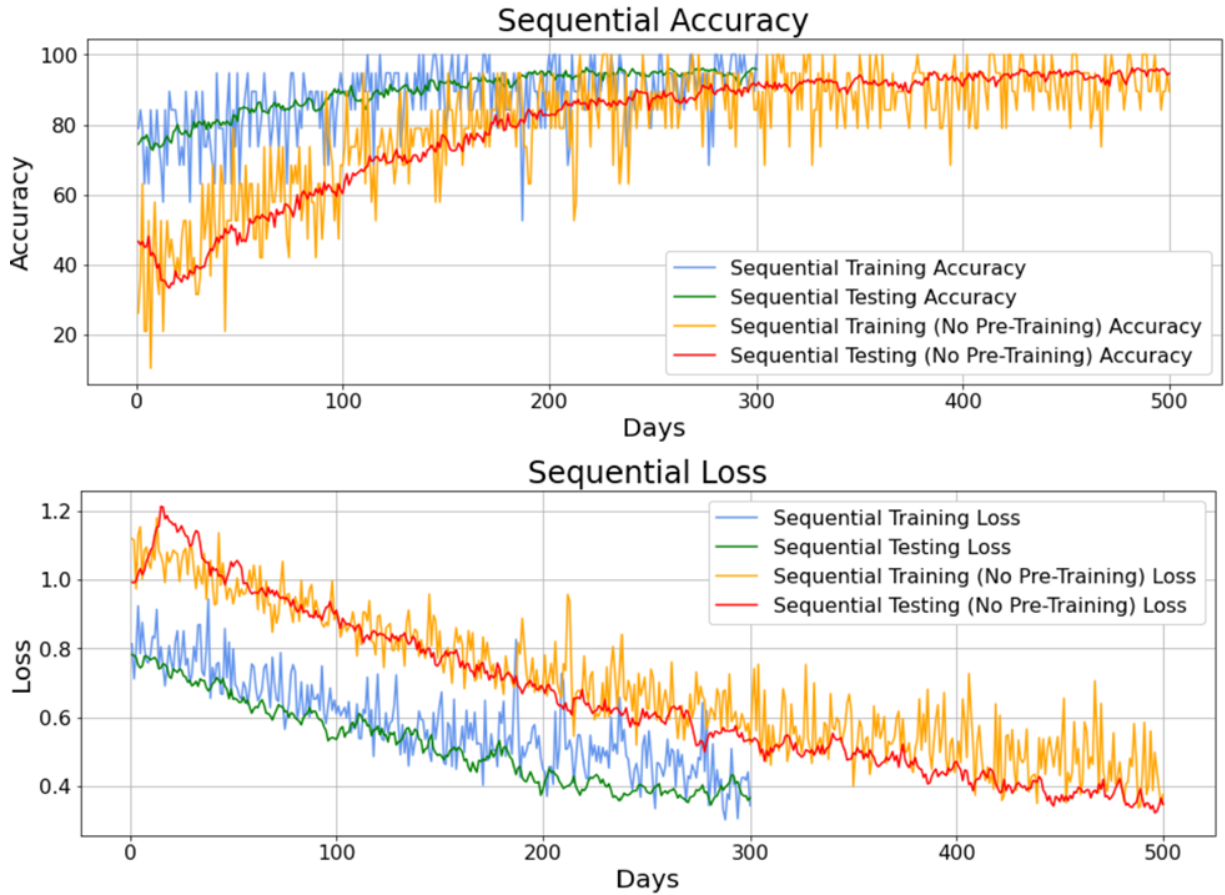

**Figure 4:** This figure shows the sequentially trained CNN with and without pre-training for one day-epoch. The accuracy of the 3-way classifier (CXR vs. Hand vs. HeadCT) as a function of "days" (top). The loss of the 3-way classifier as a function of cross-entropy loss (bottom) per "day". The training metrics are for the 20 datasets evaluated by the CNN each "day". The testing metrics are evaluated by the CNN at the end of each "day" on the set aside testing data.

### 3.2 Experiment 2: Validation Set Construction

The results shown in Figure 5 represent the findings from generating validation datasets through two methods. In Figure 5A, we can observe the sequentially trained network out to 10 "days". There is no training conducted on the first "day" due to no prior "day's" dataset to train on. Although the training and validation accuracy fluctuates over the first six "days" of sequential training, the oscillation remains compressed within a ~20% range. The validation accuracy steadily increases over the training period and finally reaches 90-100% accuracy on the final "day". There are instances of catastrophic forgetting on "days" six and eight reflected in the training and validation loss curves which see spontaneous jumps. The training loss stabilizes halfway through "day" five and the validation loss decreases steadily over the entire training period despite oscillating without an observable trend. In Figure 5B we have the results from our sequentially trained network using our second method. Here we can observe training performed on the first "day" due to N/2 images acquired from the present "day". With each subsequent "day" the network trains on 50 images and validates on 25 images without replacement. The validation accuracy fluctuates between a less compressed ~30% range each "day" compared to Experiment 2a, and the training accuracy stabilizes between 90-100% roughly halfway through the first "day" training. The validation accuracy stabilizes between 90-100% on "day" six, although it destabilizes between 80-100% over the final four "days". Catastrophic forgetting can be observed with the validation loss curve on "days" five and eight. The training loss reflects the training accuracy by stabilizing towards the end of the first "day", and the validation loss mostly fluctuates without an observable trend over the entire training period. Comparing these findings back to Experiments 1a and 1b, we demonstrate the advantages multiple day-epochs provide through sequential learning networks fully trained across fewer training "days".

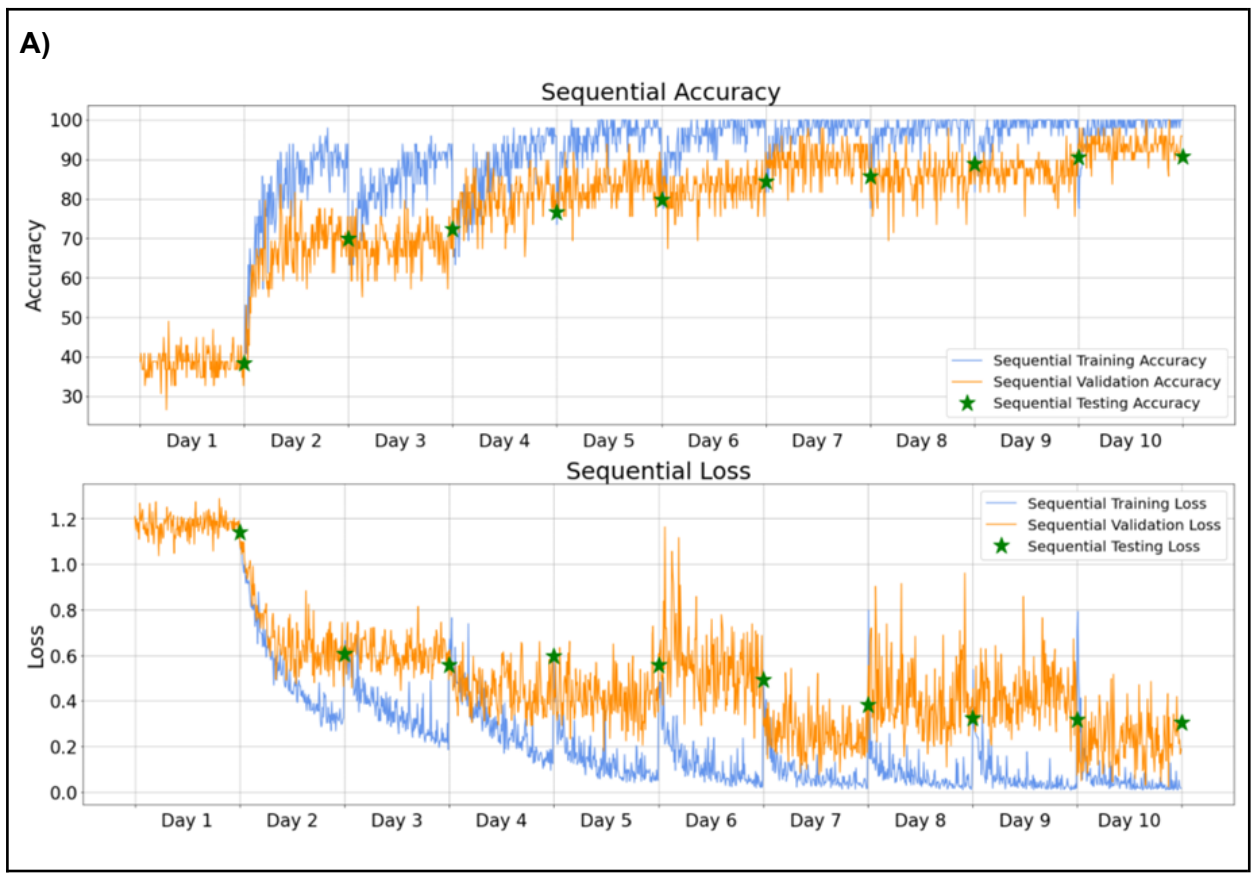

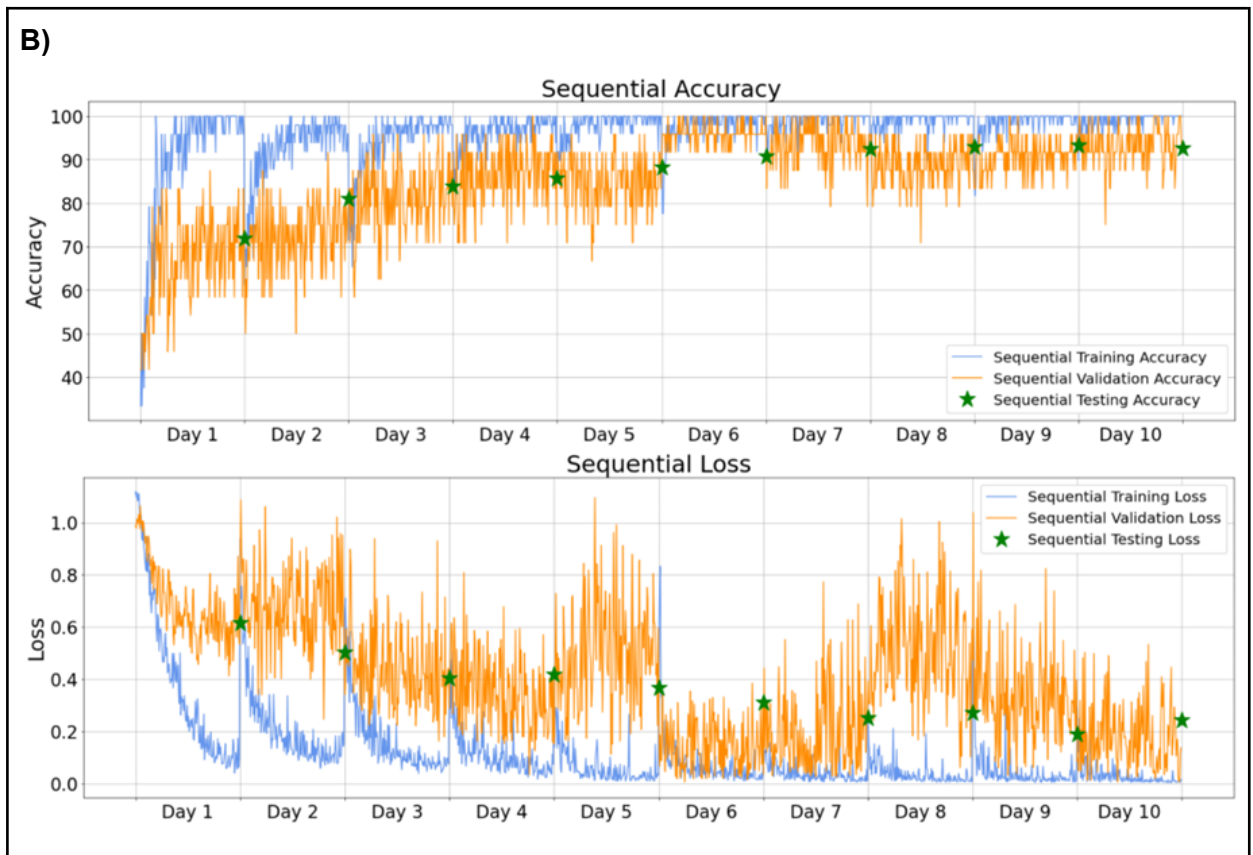

**Figure 5:** The sequentially trained CNN for multiple day-epochs. Experiment 2a **(A)** shows the accuracy of the 3-way classifier on the modified CXR data (rotated left vs. original vs. rotated right) as a function of "days" and day-epochs (top). The loss of the 3-way classifier as a function of cross-entropy loss (bottom) per day-epoch. The training metrics are for the 50 datasets evaluated by the CNN from each "previous day". We can observe how the training curves begin on the second "day". The validation metrics are evaluated by the CNN for the 50 datasets acquired in the "present day". The testing metrics are evaluated by the CNN at the end of each "day" on the set aside testing data. Experiment 2b **(B)** shows the accuracy of the 3-way classifier on the modified CXR data (rotated left vs. original vs. rotated right) as a function of "days" and day-epochs (top). The loss of the 3-way classifier as a function of cross-entropy loss (bottom) per day-epoch. The training metrics are for the 50 datasets evaluated by the CNN (N/2 from the "present day" and N/2 from the "previous day"). The validation metrics are evaluated by the CNN for the remaining N/2 datasets acquired in the "present day".

## 3.3 Experiment 3: Real-World Data

In Figure 6, we observe the results of the sequentially trained CNN with and without pre-training using the NIH Chest X-Ray dataset. Beginning with batch pre-training in Experiment 3a, our testing accuracy reached ~60% before continuing to the sequential training period where it would continue to increase. The model benefited from additional sequential training despite little improvement shown over the first 100 "days". Training accuracy fluctuates between 30-90% over the entire training period. Meanwhile testing accuracy fluctuates between 50-65% over most of the sequential training period before peaking near "day" 140 at ~69% and fluctuating past this point between 60-68%. In Experiment 3b we can observe the same CNN without pre-training over a larger period of training "days". The model trains from scratch, as observed via the testing accuracy which begins near ~32%. After acquiring images each "day" without replacement, the testing accuracy steadily increases over the first 100 "days" to reach ~60%. The final 120 "days" show the validation accuracy fluctuating between 60-68%, and the peak accuracy was achieved towards the end of the training period. The comparative results across both experiments reflect that which was shown in Experiments 1a and 1b, displaying the same benefits provided with an initial data subset though not requiring dependence on it for sequential learning. These results also align with what other users have shown with the consensus performance reaching ~65-75% classification accuracy [10,17-20].

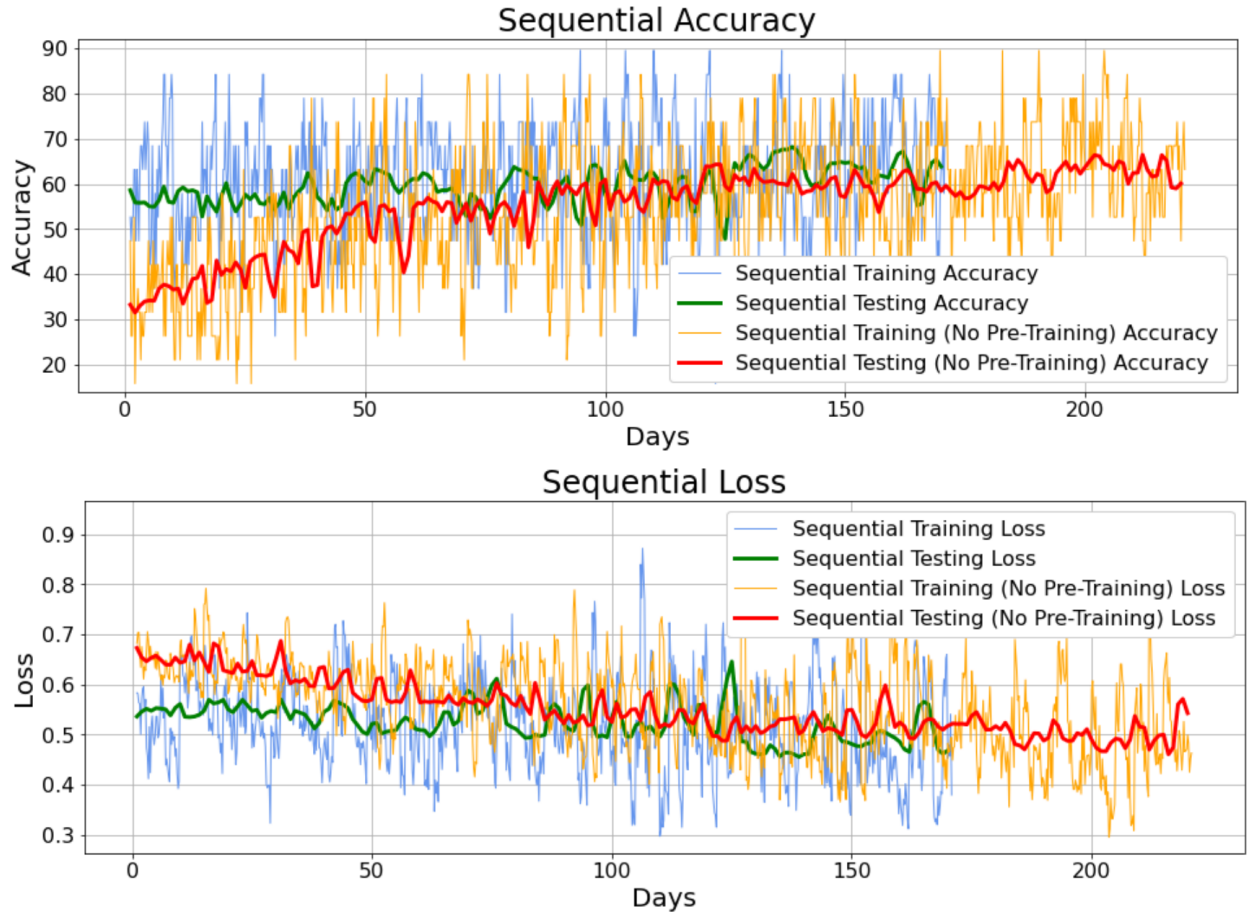

**Figure 6:** This figure shows the sequentially trained CNN with and without pre-training for one day-epoch. The accuracy of the 3-way classifier on the NIH Chest X-Ray data (effusion vs. mass vs. no finding) as a function of "days" (top). The loss of the 3-way classifier as a function of BCE with logits loss (bottom) per "day". The training metrics are for the 20 datasets evaluated by the CNN each "day". The testing metrics are evaluated by the CNN at the end of each "day" on the set aside testing data.

## Section 4: Discussion

We evaluated our sequential learning implementation on multi-classification CNNs that trained, validated, and tested on data obtained from three medical imaging datasets. Our initial experiment showed an increase in multi-classification accuracy for an untrained model over a period of learning epochs and "days". We have also shown the benefit of having an initial training data subset to pre-train a model to a desirable starting point in Experiment 1a. Although the CNN would reach desirable results sooner with pre-training data, the CNN that trains from scratch will reach the same result over a greater training period. We would next evaluate our sequential learning experimentation to reach full information extraction on acquired data by increasing the number of "day" epochs through two methods of establishing a validation dataset. Both methods would provide insight for an appropriate number of "day" epochs required for full information extraction without overfitting the CNN on the data. We did however observe signs of catastrophic forgetting for both methods, making it imperative to fine tune hyperparameters to allow the CNN to stabilize at a desirable accuracy threshold. Real-world scenarios present a similar advantage to smoothing out a sequential learning model across multiple "days". A few key differences between scans obtained over multiple "days" include scanner drift that results in signal degradation or intensity changes over time [21], different technicians acquiring data on different "days", and calibration changes for scans acquired over many weeks. We have also shown how too many day-epochs can be detrimental to a network's learning performance due to data overfitting and catastrophic forgetting. Experiments 3a and 3b evaluated a CNN with and without pre-training, respectively, with a more challenging multi-classification task. This final experiment used our "real-world" dataset, where both approaches observed overfitting with testing accuracies peaking at ~68%. Additionally, the time for training required was significantly longer than previous experiments, likely due to the NIH images having larger dimensions. Although our testing accuracies were not great with our "real-world" dataset, the results were satisfactory compared to what others have shown.

Even though our sequential learning experimentation yielded valuable findings our work contained a few key limitations, one of which was our use of public imaging datasets. Though useful for our proof-of-concept work, there remain uncertainties with more challenging datasets. The NIH Chest X-Ray dataset is a good example of a real-world dataset as it contains data captured over a 20-year period. The issue this dataset poses is differentiating subtle differences between scans as technology improves over this acquisition period. Although we implemented two unique PyTorch modes, we can explore more models that could potentially outperform ResNet50 and DenseNet121 in future work. We also aim to acquire data from a direct resource to better recognize and filter images based on scanner type, scan parameters, and diagnoses. While the Medical MNIST dataset achieved optimal results in Experiments 1a and 1b, there was greater precision tuning required for model parameters with the other two datasets due to classification difficulty increase. Moving forward we will implement optimization strategies to find the best combination of hyperparameters.

Work by other researchers has shown the ability to deal with data shifts that arise from new emergent data sources at unknown time points. With each strategy the goal remains the same: full information extraction from images in a continuous data stream. Perkonigg et al. proposed their DM strategy to continuously update parameters of a task model with novel data characteristics while sustaining diversity of the entire cohort [11]. By using data from samples stored in memory along with newer oncoming data, the model consistently extracts information from older datasets by reintroducing them during continual training, ensuring the memory is diverse and representative of all visual variations [11]. Ebrahimi et al. proposed an approach coined as adversarial continual learning (ACL) that learns a private latent space for a defined task and a shared feature space to enhance knowledge transfer and better recall previously learned tasks [22]. This approach makes the shared features less prone to forgetting due to the replay buffer mechanism, a system which takes information from samples used in prior tasks to help with better factorization and improve model performance [22]. Mittal et al. proposed a class-incremental learning technique which identified poor quality of learned representation due to effects of overfitting and loss of secondary class information [23]. They addressed validation loss in an initial base task that would drop the performance of a corresponding incremental task through regularization techniques. Such techniques included data augmentation and self-distillation which would boost the average incremental accuracy while mitigating any overtraining risks, allowing the network to retain high amounts of secondary information [23].

# Section 5: Future Directions and Conclusion

The performance of our multi-classification sequential learning technique was evaluated using public medical imaging datasets for different experiments. Particularly, our Experiment 3 with the NIH Chest X-Ray dataset was the most challenging due to its high complexity which reflects "real-world" data. Future work will include 1) acquiring medical imaging data from our own institution for pilot sequential learning studies and 2) eventually seeing our technique integrated as a research tool. Power analyses would be similarly useful to evaluate the total number of recruited data, sequential learning "days", and day-epochs required to reach a certain accuracy threshold. By recruiting N samples over M training "days", we would predict an estimate of total "days" needed to fully train a sequential learning CNN. This delicate balance of evaluating hyperparameters was more prevalent with Experiment 3. With this experiment, too few day-epochs would not allow the network enough time to learn on acquired data and too many day-epochs can negatively impact the network's performance due to overfitting. Our goal is to establish a dynamic algorithm that would adjust the day-epoch count around the number of N images acquired per "day". While our multi-classification experimentation was limited to three classes, future work will expand that number as we develop a stronger classification network. To summarize, we have shown sequential learning as a viable ML multi-classification technique capable of producing results on par with traditionally trained CNNs. Our findings demonstrate feasibility with ML and AI methods for clinically realistic scenarios where imaging data is obtained in small increments as part of routine radiologists' workflow.

## Conflict of Interests Statement
None

## Funding Sources
This research did not receive any specific grant from funding agencies in the public, commercial, or not-for-profit sectors.